# ARCHAEOASTRONOMY AND THE CHRONOLOGY OF THE TEMPLE OF JUPITER AT BAALBEK

**Giulio Magli**

Faculty of Civil Architecture, Politecnico di Milano.
Piazza Leonardo da Vinci 32, 20133 Milano, Italy
giulio.magli@polimi.it

*One of the most complex architectural feats ever conceived is the magnificent temple of Jupiter at Baalbek, Lebanon. Several issues remain unsolved about this site, and in particular the chronology and dating of the two "podia" and the true nature of the cult. We present here some hints coming from orientation and from other features of the temple, which seem to point to a unified project of both podia, originally conceived under Herod the Great.*

## 1. Introduction

The Temple of Jupiter at Baalbek (Heliopolis), Lebanon, is worldwide famous for its size and megalithic architecture (Segal 2013). In spite of this, a impressive number of problems remain unsolved about this monument, including precise dating, phases of construction, and even the true nature of the "triad" cult - Jupiter, Venus and Mercury - practiced at the place (Kropp 2009, 2010).
From the point of view of the History of Architecture, one of the problems is the absence of contemporary sources, another the desolating uniqueness of the building technique. The main features of the temple which are of interest here can be briefly described as follows. The building develops along a monumental axis comprising an hexagonal court and a huge platform with Propilea which were built by the Romans in the second century AD, as well as the nearby temples of Bacchus and Venus. The Jupiter Temple proper (or better the final phase of it) dates to 60 AD and was built on a huge basement (from now on we adopt the terminology of Kropp and Lohmann 2011 and we call it "Podium I") of squared blocks still boasting six enormous columns (almost 20 m high). At a distance of a few meters from Podium I runs a huge wall (from now on "Podium II") which surrounds as a giant U the three sides of Podium I, maintaining a strict parallelism with them. This wall was originally built without a structural connection with Podium I, as it is apparent looking at the north west side. The design is astonishing: it was conceived as the superposition of increasingly greater stones as the height increases. Big stones are in fact used at the basis, but huger stones are present in the second course, and very huge megaliths (about 500 tons each) were raised to build the third course. Finally, enormous blocks - around 4x4x20 m – were to be placed in the uppermost course; only the south-west side was however completed, putting in place the 3 famous stones which are usually called the "trilithon". At least 3 other, enormous blocks remain in the quarry some hundreds of meters to the south west, including the hugest of all, which has been recently discovered (Abdul Massih 2015).

This construction technique is very strange indeed, because the use of pre-compression in megalithic masonry – namely, putting in place huger stone blocks near the summit – is very well attested but in *polygonal* courses, where it was used to "frieze" stresses and therefore to strengthen the joints between the blocks lying beneath (see e.g. Magli 2006). Clearly such a trick does not work if the stones are placed on horizontal layers, as in Baalbek, so we have to conclude that only aesthetic reasons guided the builders. As far as dating is concerned, the temple is usually dated to Julio-claudian period (40-60 AD) and indeed a graffito, left by one of the stonemasons, brings the date of 2 August AD 60 on top of one of the column shafts, which therefore must have been almost finished by that date. However, a recent and detailed architectural analysis of Podium I has shown striking similarities - such as the use of alternating rows of headers and stretchers and of drafted-margin masonry – with Herodian sanctuaries, and in particular with what remains visible at the Temple Mount at Jerusalem, not only in general appearance, but also in proportions and measures (Kropp and Lohmann 2011). This analysis leads quite naturally the authors to conclude that a first, original phase of Podium I was built by Herodian architects. The construction can be reasonably dated around the foundation of the Augustan colony of Berytus (Beirut), which occurred in 15 BC, whose territory included Heliopolis. Since this date, Herod must have been keen in showing his attention for the Roman possessions in the area.

It is the aim of the present paper to analyze the Temple of Jupiter starting from the point of view of modern Archaeoastronomy (see e.g. Magli 2015), and therefore to study the building within the sky landscape it is immersed in and in the context of similar monuments. As we shall see, our results do support the idea that the original project of Podium I was conceived under Herod the Great. We also propose, however, that the Herodian project already included also the construction of Podium II.

## 2. Archaeoastronomy at Baalbek

The astronomical orientations of temples in the cultures of the Mediterranean has been widely investigated (see e.g. Boutsikas 2009 for Greek temples and Belmonte, Shaltout, and Fekri 2009 for Egyptian temples) but – at least to the best of the author's knowledge - the orientation of the Baalbek's temple of Jupiter has not been studied. This orientation can be determined as follows. The area is quite well covered by satellite imagery and from satellite data (Google Earth and Bing) both the azimuth and the horizon height (from inside looking out) can be determined with good approximation (say within 1/2 degree for both). The results are azimuth 75° 30', horizon height 5°. Using the program getdec (kindly provided by Clive Ruggles) which takes into account refraction, at the latitude of Baalbek these data yield a declination of ~14° 44'. This declination is within the solar range: the sun therefore rises in alignment with the temple twice a year, around May 1 and August 12 (Gregorian, but of course up to the second century AD the difference with the Julian was of less than one day). These dates do not seem to be of special significance for the cult, and this may be seen as a confirmation of already existing doubts (Kropp 2010) on the true solar character of the "Heliopolitan" Jupiter. In fact, solar temples (for instance in Egypt) have been usually oriented to the true east (Magli 2013). This holds, for instance, for the sun temples of the 5th dynasty and, with all probability, for the precinct of Egyptian Heliopolis, whose connection with Baalbek-Heliopolis has been claimed several times. In case of complex deities with solar connotations, such as Amun Ra at Karnak, the orientation was anyhow to a special day of the year, the winter solstice.

The solar dates do not appear of special significance also in the Roman calendar (the foundation of Rome, 21 April, being too distant) and in the Hebrew luni-solar calendar (the Passover feast does never stretch up to the end of April). Further, the orientation at Baalbek can be compared to those of the other three main temples of Zeus in the region, namely Kanawat, Damascus, and Gerash. The results are:

1) Kanawat points almost to true north (azimuth 4°, horizon flat or nearly flat); the building is very clealry directed to the center of Philippopolis, which lies some 10 Kms to the north and was planned a few decades later, probably using the temple as a landmark

2) Damascus (azimuth 85°, horizon nearly flat) has a generic solar orientation which is conformal to the orthogonal grid of the town,

3) The huge sanctuary of Zeus at Gerash has azimuth 58°, horizon height 3° 15', yielding an impressive declination of +28 31 (increasing at 29° for the geocentric lunar declination), and may thus have been oriented to the maximal northern standstill of the Moon, a possibility that certainly deserves further research.

This, as far as what concerns us here, there is no solar pattern in the orientation of these temples as well, and orientation appears to be specific for each temple. What about Baalbek? The azimuth is not governed by the topography either – the huge platform was oriented exactly where the builders wanted it to be, without any geological constraint - and so we are led to investigate about a possible stellar alignment. Analyzing the sky *at Herod's times* it can be seen that a quite important celestial object was rising in alignment with the temple: the Pleiades.

The Pleiades are an asterism, not a single star; however, their apparent dimension in the sky is very small and they can be considered (and were considered in antiquity, since Hesiod times in the Greek world) as a single entity, although seven stars can be distinguished with the naked eye. To fix ideas we can consider the declination of the Pleiades as a whole asterism to be between 15° 30' and 16° (the star Alcyone in 15 BC had a declination of 15° 50'). The agreement is therefore good (~1°), and the horizon height assures that the asterism was really visible (faint stars are not visible until they are at an height at least comparable to their magnitude in degrees).

Clearly, there is the distinct possibility that this alignment occurs just by chance. However, interest for the Pleiades is well documented in the Greek religion (for instance, the role of this asterism has been shown to be fundamental for the rites of Artemis Orthia sanctuary in Sparta, Boutsikas and Ruggles 2011; see also Boutsikas and Hannah 2012 for the role of the Hyades at Athens' Acropolis). Is it possible to associate the Pleiades with the Heliopolitan Jupiter? The pre-Roman history of the God is uncertain, but the iconography is well attested from the Roman period and from the unique written source we have on Baalbek, the Saturnalia dialogues of the fifth-century author Macrobius. The cult image represents a young, unbearded Jupiter, bringing a huge vase-shaped top hat (*Kalathos*). The God usually brings grain ears and a whip, and is accompanied by two walking bulls. The Heliopolitan Jupiter thus had very clear attributes of a God of fertility: the *Kalathos* and the grain ears. In this connection, the orientation to the Pleiades becomes more understandable. Indeed already in Hesiod calendar (8th century BC) the harvest time of the cereals was indicated by the heliacal rising of Pleiades which occurred in the first week of May. Taking into account the shift in time but also in latitude, this date does not change substantially in Baalbek, where the Heliacal rising of the Pleiades occurred around 5 May in 15 BC. Interestingly enough, therefore, the alignment of the temple (roughly) individuated both the heliacal rising of the Pleiades and that of the sun which occurred a few minutes later. Before in the same nights, one could see in the very same direction the rising of the constellation Aries (the declination of the star Hamal was ~13°) a constellation associated with spring and renewal as well (a direct association of Aries with Zeus probably existed, but in the case of Zeus-Ammon, the horned God of Egyptian origin which is not documented at Baalbek).

## 3. Discussion

Due to the phenomenon called precession, the declination of each star slowly and continuously changes. In the case of the Pleiades, the alignment at Baalbek worsened with time: for instance in 60 AD Alcyone had a declination of 16° 15'. There are a few examples of buildings whose axis was slightly re-planned to follow the rising of a star, and so – eventually - a slight deviation of the axes of Podium II with respect to Podium I might have provided a clue to its dating, but this is not the case. The strict parallelism of the two leads us therefore to reconsider the problem of Podium II.

In non-scholarly publications (see e.g. Hancock 2015) prevails the idea of a great antiquity. However, and letting of course alone any "lost civilization" theory, it is easy to see that *any* predating (say, also of a few decades) of the megalithic wall with respect Podium I would have led the architect of the latter *to use it*. Only a fool could construct ex-novo a huge basement, oriented in the same way and accurately placed just a few meters inside the existing wall, without taking advantage of it as a ready-to-go, tremendously stable and affordable retain structure.

So far so good for the "Podium II predates Podium I" theory. The reverse theory implies that the style of Podium I was not acceptable to the Roman standards and therefore in Julio-Claudian times the Romans opted for the enlargement of the building (Kropp and Lohmann 2011). However, any Roman architect willing (for "stylistic" reasons which, at least to the present author, seem quite weak on their own) to enlarge Podium I up to the dimensions of Podium II, would have *expanded* the existing basement up to the desired dimension by adding courses of stone blocks. Constructing ex-novo a self-standing, gargantuan megalithic wall with the idea of filling the gap between the two *later on* is a behavior almost as illogical as the one implied by the inverse chronology.[1]

---

[1] The presence of Roman sketch engravings of the temple pediment on one of the blocks of the Trilithon has been claimed as a proof of contemporaneity. Of course it is not: the Roman architects were used to sketch their projects on pre-existing monuments, for instance on the paved floor in front of Augustus mausoleum a precise drawing of Hadrian's Pantheon pediment can be seen. Another proof should be that in the lower course a piece of column drum was used instead of a block; however - if it was not a Arab repair - the column might belong to the Herodian temple

Excluding the impossible, what remains must be the truth: I propose that the absence of structural connection – and simultaneously the strict parallelism -  between the Podia can be explained only if two structures were planned together. In accordance with the dating we are supporting for Podium I, therefore, the whole project would have been conceived by Herod the Great. In this respect it should be noticed that strict architectural analogies with the Herodian architecture at the Temple Mount do hold *also* for Podium II. In fact huge foundations, made of gigantic stone blocks, have been unearthed in the tunnel excavated along the western wall of the Mount (Ritmeyer 1992, Bahat 1994). These blocks show beyond any reasonable doubt that megalithic masonry was in the mind, and in the abilities, of Herodian stonemasons: the hugest known of the Herodian blocks in Jerusalem is indeed 13.7x3.2x4.2 mt, weighing about 570 tons. Further also there, as in Baalbek, the hugest stones are not set at the lowest courses, which are instead made of smaller blocks.

Why did Herod's architects built Podium II? A possibility is that they wanted to form a U-shaped gallery encircling the sides of the temple, similar to the later, huge gallery which still today encircles the front platform. The function of the back gallery might have been related to the cult, perhaps to exploit oracular rites. The Gargantuan project remained unfinished and, in particular, the builders did not succeed in terminating the exterior side walls with the transport of the missing megaliths, so the construction of the vaults did not begin. As a consequence, the megalithic wall remained as a sort of (at this point really anti-esthetic) curtain and this explains why in the Julio-Claudian stage it was decided to fill it with blocks of stone. The temple we can see today is the final result of the Arab conversion of the building in a fortress, with walls built with second-use blocks.